**Exact solution of the two-dimensional (2D) Ising model at an external magnetic field**

Zhidong Zhang

Shenyang National Laboratory for Materials Science, Institute of Metal Research, Chinese Academy of Sciences, 72 Wenhua Road, Shenyang, 110016, P.R. China

**Abstract**

The exact solution of the two-dimensional (2D) Ising model at an external magnetic field is derived by a modified Clifford algebraic approach. At first, the transfer matrices are analyzed in three representations, *i.e.,* Clifford algebraic representation, transfer tensor representation and schematic representation, to inspect nonlocal effects in this many-body interacting system. It is ensured that nontrivial topological structures exist in this system, which is analogous to (but different with) those in the three-dimensional (3D) Ising model at zero magnetic field. Therefore, the approaches developed for the 3D Ising models are modified to be appropriable for solving analytically the solution of the 2D Ising model at a magnetic field. An additional rotation, serving as a topological Lorentz transformation, is applied for dealing with the topological problems in the present system. The rotation angle for the transformation is determined by Yang-Baxter relations and a subsequent average of rotation angles treating the linear change of the topological actions. Application of a magnetic field increases the magnetization, shifting the critical point to higher temperatures. At the temperature above the critical point, the magnetization keeps zero until a critical field $H_c$ at which it jumps rapidly. However, when the magnetic field is renormalized by the critical field with respect to the relative temperature $\Delta T = (T - T_c)$, the magnetization processes above the critical point have the same character as that in the critical point. The nature of such a jump in the nonzero magnetic field is neither a first-order magnetization process nor a phase transition. The partition function and the magnetization obtained are helpful for understanding the physical properties, in particular, the magnetization processes of the 2D magnetic materials.

The corresponding author: Z.D. Zhang, e-mail address: zdzhang@imr.ac.cn

## 1.Introduction

The Ising model is one of the simplest physical models describing many-body spin (or particle) interactions [1]. Kramers and Wannier [2] employed the duality between the interactions $K$ and $K^*$ to fix the critical point for the two-dimensional (2D) Iisng model. Onsager [3] solved exactly the partition function, the free energy and the specific hear of the 2D Ising model. Kaufman [4] developed a spinor representation to simplify the procedure for deriving the exact solution. Yang [5] derived the exact solution of the spontaneous magnetization of the 2D Ising model by a perturbation procedure. Kac and Ward [6] developed a combinatorial solution of the 2D Ising model.

Newell and Montroll [7] analyzed the theory of Ising models and pointed out that the difficulties for exactly solving the three-dimensional (3D) Ising model at zero magnetic field, the 2D Ising model with the next nearest interactions at zero magnetic field, and the 2D Ising model without the next nearest interactions at the presence of an external magnetic field are topological. One meets serious hinders as one attempts to apply the algebraic method used for the 2D model at the absence of a magnetic field to the 3D Ising model or to the 2D Ising model with a magnetic field or the next nearest interactions. The operators of interest generate a much large Lie algebra that it would be of little value, while nontrivial topological structures emerge. The procedures developed by Onsager [3], Kaufman [4], Kac and Ward [6] cannot be generated directly to be applicable for these two problems, which introduce some troubles in topology by counting closed graphs. However, the topological problems are different for the three models: In three dimensions, one encounters polygons with knots; The magnetic field problem involves a much more complicated procedure, counting not only of the number of bonds in the polygon but also the area. The peculiar topological property is that a polygon in three dimensions does not divide the space into an "inside and outside" [7]. Any approaches based on only local environments cannot be exact for the 3D Ising model, or for the 2D Ising model at a magnetic field or with the next nearest interactions.

In order to solve the ferromagnetic 3D Ising model, the present author proposed two conjectures in [8], investigated its mathematical structure in [9], then proved rigorously the two conjectures by a Clifford algebraic approach in collaboration with Suzuki and March [10], and further by a method of Riemann-Hilbert problem in collaboration with Suzuki [11,12]. The critical exponents were derived exactly to be $\alpha = 0$, $\beta = 3/8$, $\gamma = 5/4$, $\delta = 13/3$, $\eta = 1/8$ and $\nu = 2/3$ [8]. Furthermore, the exact solutions of the two-dimensional (2D) Ising model with a transverse field [13] and the 3D $Z_2$ lattice gauge theory [14] were derived by the equivalence/duality between these models. Based on these results, topological quantum statistical mechanics and topological quantum field theories were investigated systematically [15]. The experimental data confirm the existence of the 3D Ising universality class [16,17], which affirm the validity of the exact solutions of the 3D Ising models. The Monte Carlo simulations [18,19] on the critical exponents of the 3D Ising model, which were obtained by taking into account the nontrivial topological contributions of spin chains, agree well with our exact solutions. In addition, guided by a better understanding on the nontrivial topological structures in the 3D Ising models [8-12], the present author determined the lower bound of computational complexity of several NP-complete problems, such as spin-glass 3D Ising models [20], Boolean satisfiability problems [21], knapsack problems [22] and traveling salesman problems [23]. Furthermore, the 2D Ising model with the next nearest interactions at zero magnetic field was studied analytically [24].

The motivations of the present work are as follows: The low-dimensional magnetic materials have attracted intensive interests in recent decades. The exact solution of the 2D Ising model at an external magnetic field is a long-standing unsolved problem in physics, which is in the same difficulty level as the exact solution of the ferromagnetic 3D Ising model at zero magnetic field. The approaches developed for solving exactly the 3D Ising model at the absence of a magnetic field may guide the route to derive the exact solution of the 2D Ising model at a magnetic field. The exact solution is extremely important for inspecting physical properties and specially magnetization processes of the 2D Ising model as well as the 2D magnetic materials.

This paper is arranged along the following line of presentation: In Section 2, the model is set up and the transfer matrices are described respectively by a Clifford algebraic representation, a transfer tensor representation and a schematic representation, to inspect nontrivial topological structures. In Section 3, the Clifford algebraic approach developed previously for solving the 3D Ising model at the absence of a magnetic field [10] is modified for solving the present system. In Section 4, the eigenvalues, the partition function and the magnetization of the 2D Ising model at a magnetic field are derived explicitly by a modified Clifford algebraic approach. In Section 5, we discuss the dimensionality, rotation angles and topological phases. Section 6 is for conclusions.

## 2. Model and transfer matrices

The Hamiltonian of the 2D Ising model at an external magnetic field is written as:

$$\hat{\mathrm{H}} = -\sum_{<i,j>}^{m,n} \left[ J_1 s_{i,j} s_{i+1,j} + J_2 s_{i,j} s_{i,j+1} + h s_{i,j} \right]$$

(1)

Here every Ising spin $s_{i,j}$ located on a rectangular lattice with the lattice size $N = mn$ takes two values +1 and -1 for spin up and spin down. The numbers $(i, j)$, running from (1, 1) to $(m, n)$, denote lattice points along two crystallographic directions. Only are the nearest neighboring interactions $J_1$ and $J_2$ between spins considered, which are all ferromagnetic. One may set $J_1 \geq J_2$ without loss of generality. The external magnetic field $h$ acts on every spin in all the 2D lattice sites.

2.1. Clifford algebraic representation

The partition function $Z$ of the 2D Ising model at a magnetic field is expressed as follows:

$$Z = (2sinh2K_1)^{\frac{n}{2}} \cdot \mathrm{trace}(\boldsymbol{V})^m \equiv (2sinh2K_1)^{\frac{n}{2}} \cdot \sum_{i=1}^{2^n} \lambda_i^m$$

(2)

with the transfer matrix $\boldsymbol{V} = \boldsymbol{V_3} \boldsymbol{V_2} \boldsymbol{V_1}$ as:

$$\boldsymbol{V_3} = \prod_{j=1}^{n} exp\left[ -iH \left( \prod_{k=1}^{j-1} i\, \Gamma_{2k-1} \Gamma_{2k} \right) \Gamma_{2j-1} \right]$$

(3)

$$\boldsymbol{V_2} = \prod_{j=1}^{n} exp\left[ -iK_2 \Gamma_{2j} \Gamma_{2j+1} \right]$$

(4)

$$\boldsymbol{V_1} = \prod_{j=1}^{n} \exp\left[ iK_1^* \Gamma_{2j-1} \Gamma_{2j} \right]$$

(5)

It is convenient to introduce variables $K_1 \equiv J_1/(k_B T)$ and $K_2 \equiv J_2/(k_B T)$ instead of interactions $J_1$ and $J_2$, while variable $H \equiv h/(k_B T)$ instead of the magnetic field $h$. The Kramers-Wannier relation is used to define the dual interaction $K_1^*$ [2]:

$$K_1^* = \frac{1}{2} ln(cothK_1) = tanh^{-1}(e^{-2K_1}) \tag{6}$$

The generators of Clifford algebra are written as:

$$\Gamma_{2j-1} \equiv P_j = \mathrm{C} \otimes \mathrm{C} \otimes \ldots \otimes \mathrm{C} \otimes \mathrm{s}' \otimes \mathrm{I} \otimes \ldots \otimes \mathrm{I} \quad (j-1 \ \ times \ \ C) \tag{7}$$

$$\Gamma_{2j} \equiv Q_j = \mathrm{C} \otimes \mathrm{C} \otimes \ldots \otimes \mathrm{C} \otimes (-is'') \otimes \mathrm{I} \otimes \ldots \otimes \mathrm{I} \quad (j-1 \ \ times \ \ C) \tag{8}$$

The $\Gamma_{2j-1}$ and $\Gamma_{2j}$ (and also $P_j$ and $Q_j$) matrices are referred to the Onsager-Kaufman-Zhang notations [3,4,8-12]. $\mathrm{s}'' = \begin{bmatrix} 0 & -1 \\ 1 & 0 \end{bmatrix}$ $(= -i\sigma_2)$, $\mathrm{s}' = \begin{bmatrix} 1 & 0 \\ 0 & -1 \end{bmatrix}$ $(= \sigma_3)$, $C = \begin{bmatrix} 0 & 1 \\ 1 & 0 \end{bmatrix}$ $(= \sigma_1)$, where $\sigma_j$ ($j$ = 1,2,3) are Pauli matrices, while $\boldsymbol{I}$ is the unit matrix. In the transfer matrices, the boundary factor $\boldsymbol{U}$ in Kaufman's paper [4] is neglected, since it splits the space into two subspaces, and in the thermodynamic limit the surface to volume ration vanishes for an infinite system according to the Bogoliubov inequality [10].

For the 2D Ising model with a magnetic field, the Jordan-Wigner transform [25,26] gives $\sum_j \sigma_j^x$ in the transfer matrix $\boldsymbol{V_3}$ for the magnetic field with high-order terms.

$$\left[ \prod_{k=1}^{j-1} i\, \Gamma_{2k-1} \Gamma_{2k} \right] \Gamma_{2j-1} = \mathrm{I} \otimes \mathrm{I} \otimes \mathrm{I} \otimes \mathrm{I} \otimes \cdots \otimes \mathrm{I} \otimes \mathrm{s}' \tag{9}$$

The action of the magnetic field is to place the Pauli matrix $\mathrm{s}' = \sigma_3$ in the $j$-th position in the direct products of unit matrices $\boldsymbol{I}$ in the transfer matrix $\boldsymbol{V_3}$. The nonlinear terms in the transfer matrix $\boldsymbol{V_3}$ cause the topological problems including nonlocality, nonlinearity, non-commutative and non-Gaussian. Similar to the 3D case [10], these problems are roots of difficulties hindering exactly solving the problems. The difference between the transfer matrices $\boldsymbol{V_3}$ in the 3D Ising model at zero magnetic field and the 2D Ising model at a magnetic field is elucidated as follows: In the 3D case without a magnetic field, the internal factors in $\boldsymbol{V_3}$ remain the same for all the lattice pints $j$ = 1, 2, …, $nl$. The number of $\Gamma_{2j}$ matrices in $\boldsymbol{V_3}$ is always equal to 2$n$. In the 2D case with a magnetic field, the nonlinear factors in $\boldsymbol{V_3}$ alter with changing the lattice point $j$ = 1, 2, …, $n$. The number of $\Gamma_{2j}$ matrices in $\boldsymbol{V_3}$ equals to the odd numbers 1, 3, 5, …, 2$j$-1, with changing $j$ through the lattice. Such an algebraic difference results in the difference between the nontrivial topological structures in the two models. On one hand, the 2D Ising model at a magnetic field exhibits a topological problem analogous to the 3D Ising model at zero magnetic field. We may employ the Clifford algebraic approach developed in [10] to solve exactly the 2D Ising model at a magnetic field. An additional rotation is added to trivialize the nontrivial topological structures to be diagonalizable, while topological phases are generalized on eigenvectors and eigenvalues. On the other hand, the difference between the algebraic formulas for the transfer matrices and subsequently the topological structures of the two models recommends the modifications of the Clifford algebraic approach.

2.2. Transfer tensor representation

The partition function $Z$ of the 2D Ising model at a magnetic field can be represented also in the transfer tensor forms:

$$Z = (2sinh2K_1)^{\frac{n}{2}} \cdot \text{trace}(\boldsymbol{T})^m \tag{10}$$

with the transfer tensor $\boldsymbol{T}$ as:

$$\boldsymbol{T} = \prod_{j=1}^{n} exp\left[T_{uvw}^{(j)}\right] \tag{11}$$

The basic elements of the transfer tensor for the ferromagnetic 2D Ising model in a magnetic field can be expressed by the transfer tensor representation:

$$T_{uvw} = \langle s_{i,j}, s_{i+1,j}, s_{i,j+1} \left| e^{-\hat{H}/k_B T} \right| s_{i,j}, s_{i+1,j}, s_{i,j+1} \rangle \tag{12}$$

where a spin $s_{i,j}$ interacts with the two nearest neighboring spins $s_{i+1,j}$, and $s_{i,j+1}$, along two crystallographic directions in a unit cell. All possible combinations of three Ising spins with values ± 1 give eight configurations, which can be described by a three-order tensor $T_{uvw}$ with *u*, *v*, *w* = 1, 2. as follows:

$$T_{111} = \langle + + + \left| e^{-\hat{H}/k_B T} \right| + + + \rangle = e^{K_1^* + K_2 + H}, \tag{13}$$

$$T_{211} = \langle - + + \left| e^{-\hat{H}/k_B T} \right| - + + \rangle = e^{-K_1^* - K_2 - H}, \tag{14}$$

$$T_{121} = \langle + - + \left| e^{-\hat{H}/k_B T} \right| + - + \rangle = e^{-K_1^* + K_2 + H}, \tag{15}$$

$$T_{221} = \langle - - + \left| e^{-\hat{H}/k_B T} \right| - - + \rangle = e^{K_1^* - K_2 - H}, \tag{16}$$

$$T_{112} = \langle + + - \left| e^{-\hat{H}/k_B T} \right| + + - \rangle = e^{K_1^* - K_2 + H}, \tag{17}$$

$$T_{212} = \langle - + - \left| e^{-\hat{H}/k_B T} \right| - + - \rangle = e^{-K_1^* + K_2 - H}, \tag{18}$$

$$T_{122} = \langle + - - \left| e^{-\hat{H}/k_B T} \right| + - - \rangle = e^{-K_1^* - K_2 + H}, \tag{19}$$

$$T_{222} = \langle - - - \left| e^{-\hat{H}/k_B T} \right| - - - \rangle = e^{K_1^* + K_2 - H}, \tag{20}$$

The transfer tensors of the whole system can be represented by the direct products of the transfer tensors for all the spins in the 2D Ising model.

On the other hand, in the three dimensions, a spin $s_{i,j,k}$ interacts with the three nearest neighboring spins $s_{i+1,j,k}$, $s_{i,j+1,k}$, and $s_{i,j,k+1}$ along three crystallographic directions in a unit cell. The basic elements of the transfer tensor for the ferromagnetic 3D Ising model at the absence of a magnetic field can be expressed by a four-order tensor $T_{uvwt}$ with $u$, $v$, $w$, $t$ = 1, 2. The sixteen elements in the four-order tensor correspond to sixteen configurations, for all the possible combinations of four Ising spins with values ± 1. Comparing the three-order transfer tensor $T_{uvw}$ with the four-order transfer tensor $T_{uvwt}$, we find that the former is exactly the same as the latter with $t$ = 1, but the magnetic field $H$ replaces the third interaction $K_3$. The sign of the magnetic field $H$ in the former differs with that of the third interaction $K_3$ in the latter with $t$ = 2. This indicates that the two models have similar topological effects, but with some differences. However, it should be emphasized that the three-order tensor $T_{uvw}$ occupies the half space of the four-order transfer tensor $T_{uvwt}$. Therefore, we can employ the approaches developed for the 3D Ising model to solve the 2D Ising model with a magnetic field, however, with some modifications.

The three-order transfer tensor $T_{uvw}$ can be seen as a cubic matrix, while the four-order transfer tensor $T_{uvwt}$ is seen as a hypercubic matrix. In principle, one can develop a diagonalization process for a cubic (or hypercubic) matrix. The advantage of the transfer tensor representation is that it illustrates concisely all the terms of the Hamiltonian in a stereoscopic form. The nontrivial topological structures are hidden in the cubic (or hypercubic) formulates, since the cubic (or hypercubic) matrix has the 3D character that fits well with the 3D character of the basic elements for the 2D Ising model with a magnetic field (or the 3D Ising model). The disadvantage of the transfer tensor representation is that it cannot derive directly the desired solution, because the Lie algebra is so small that the number of the body-diagonalization elements are not enough for eigenvalues of the system. Meanwhile, the corresponding eigenvectors are not large enough for representing the Hilbert space of the system. Nevertheless, the transfer tensor representation gives some implications on the procedure for solving the problem. For instance, according to the transfer tensor representation, four groups of plane rotations along four axes $K_1^*$, $K_2$, $H$ and $H'$ exist in the system, where the rotation along the $H'$ axis is an additional one analogous to the additional rotation $K'''$ for the 3D Ising model.

2.3. Schematic representation

In order to illustrate the topological structures of the 2D Ising model with a magnetic field, we utilize the schematic representation to demonstrate the transformation from one topological state to others, which keeps the equivalence of the free energy of the system. Figure 1 shows schematically a 2D Ising model on a 6×6 lattice with a magnetic field applied at each lattice point. In previous work [20-23] for determining the lower bound of the computational complexity of NP-complete problems, an absolute minimum core (AMC) model in the spin-glass 3D Ising model was defined as a spin-glass 2D Ising model interacting with its nearest neighboring plane. The 2D Ising model with a magnetic field has the same structure as the AMC model, but now the interactions within the 2D lattice are all ferromagnetic and the magnetic field is acting on a spin at each lattice point. The action of the magnetic field effectively likes an interaction along the third crystallographic direction, but with a different character: The interaction is between two spins, depending on the configuration combinations of the two spins, whereas the magnetic field is acting on a spin, depending only on its configurations. Nevertheless, this similarity reveals that the topology of the 2D Ising model with a magnetic field is the same as that of the AMC model that is seen as the basic element of the 3D Ising models. This fact strongly suggests that the approaches developed for the 3D Ising model can be employed for solving the 2D Ising model with a magnetic field.

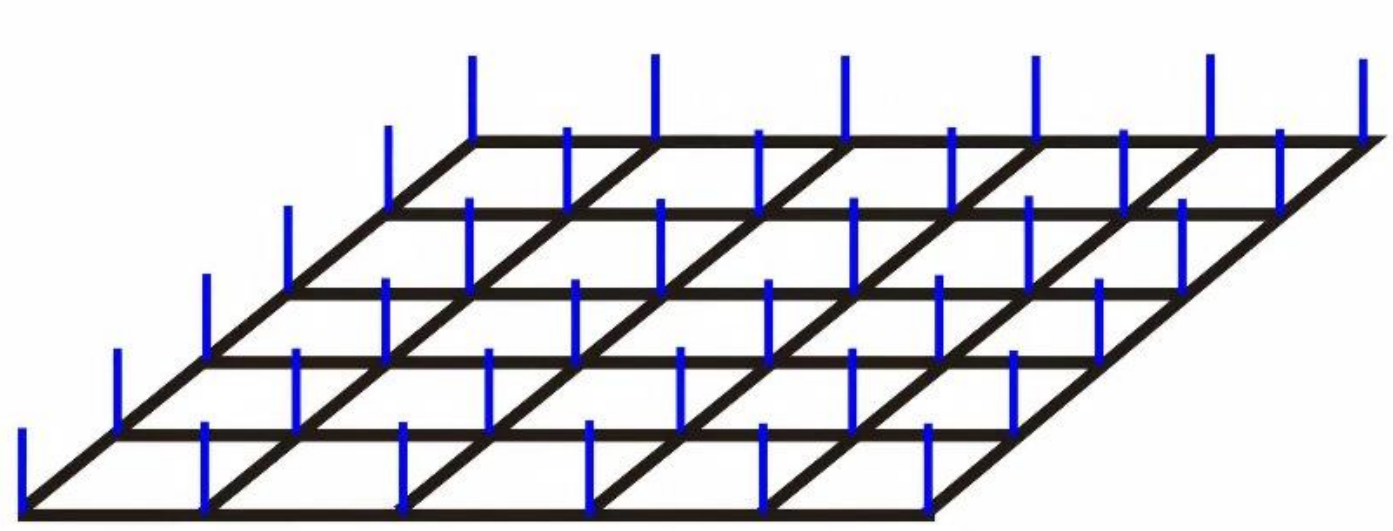

Figure 1. Illustration of a 2D Ising model on a 6×6 lattice (black solid lines), with a magnetic field (blue solid lines) applied at each lattice point.

Figure 2 illustrates a topological structure of a 2D Ising model on a 6×6 lattice, in which all the magnetic fields extend from every lattice point to the intersection at infinite. The extension of the magnetic fields to infinite does not alter the topology of the model, while keeping the physical properties unchanged. Figure 3 presents a scheme of a 2D Ising model on a 6×6 lattice with an additional row as a virtual boundary, where the magnetic field at every lattice point connects correspondingly with every virtual lattice point at the boundary. Notice that in Figures 1-3, as an example, the 2D lattice with finite lattice points are illustrated. In the thermodynamic limit, the number $N = nm$ of the lattice points approaches infinite (*i.e.*, $n\rightarrow \infty$, $m\rightarrow \infty$, $N \rightarrow \infty$). The intersection at infinite is equivalent to all the virtual lattice points at the boundary infinite far. The systems demonstrated in Figures 1-3 can be transformed from one to another without altering the physical properties of the system. Topologically, they are equivalent. The nontrivial topological structures can be illustrated more clearly in Figure 3, which are consistent with those hidden in the Clifford algebraic formulas (Eqs. (2)-(5)) for the transfer matrices.

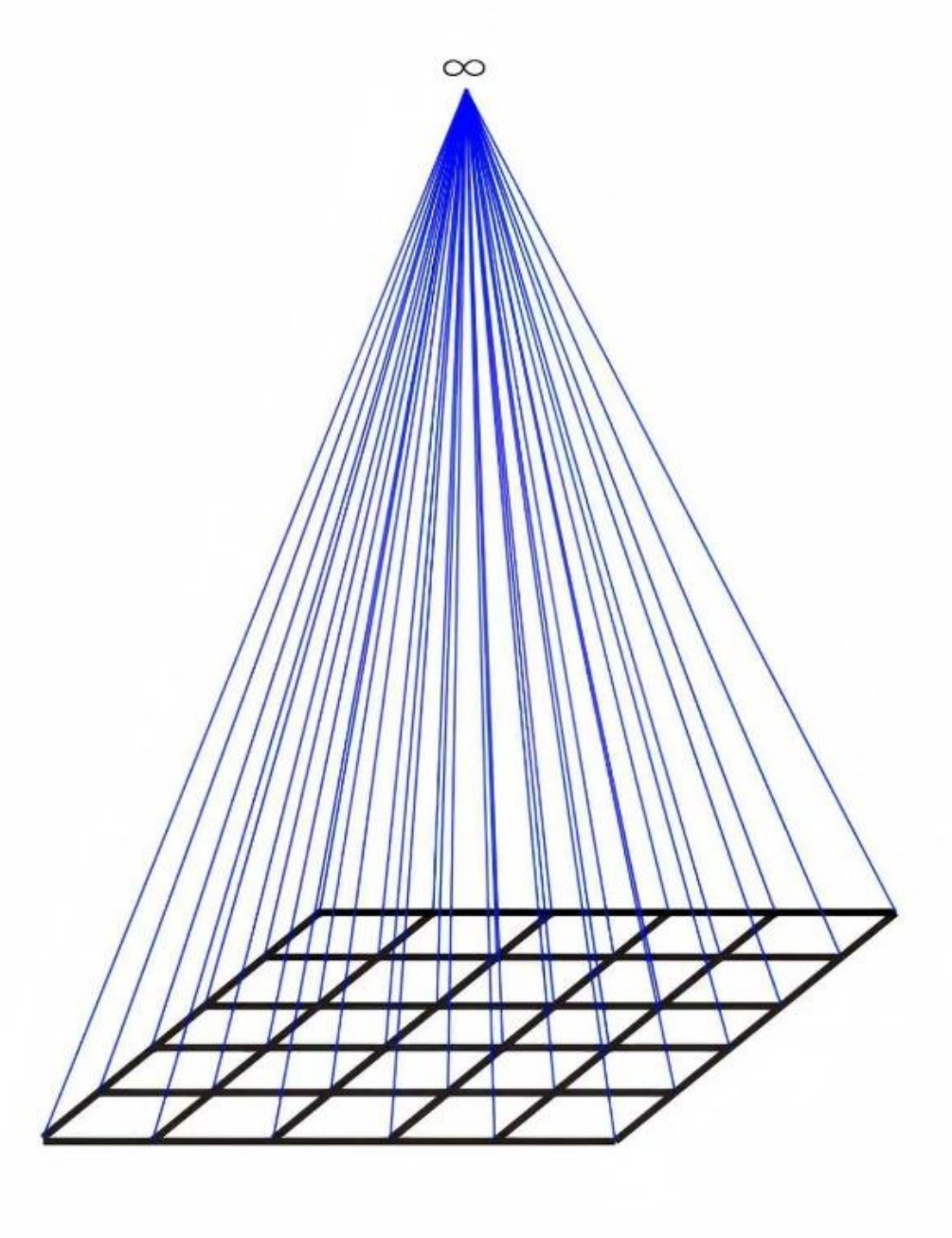

Figure 2. Illustration of a 2D Ising model on a 6×6 lattice (black solid lines), in which all the magnetic fields (blue solid lines) extend from every lattice point to the intersection at infinite.

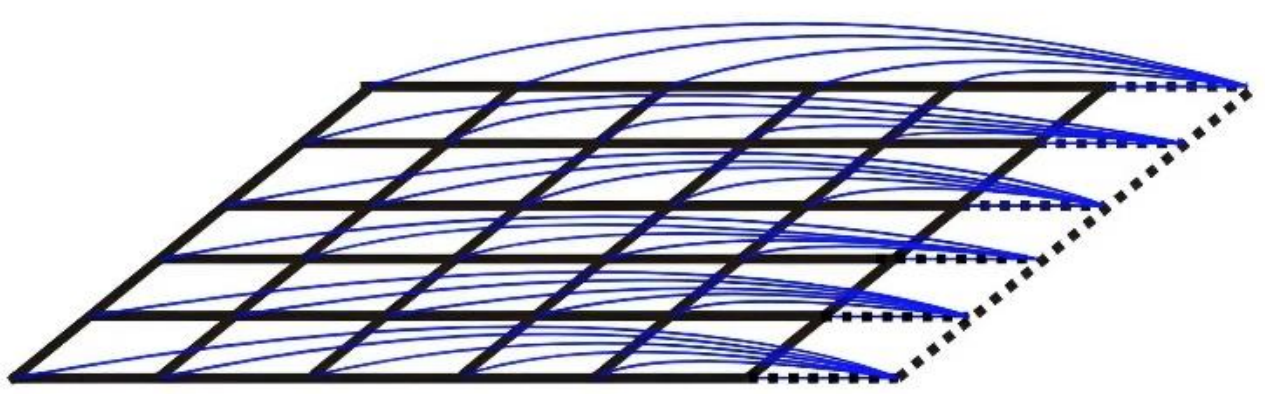

Figure 3. Illustration of a 2D Ising model on a 6×6 lattice (black solid lines) with an additional row (black dash lines) as a virtual boundary, where the magnetic fields (blue solid lines) at every lattice point connect with the virtual lattice points at the boundary.

It is worth inspecting the difference between the nontrivial topological structures between the 2D Ising model with a magnetic field and the 3D Ising model at zero magnetic field. In the 3D Ising model, the interaction along the third crystallographic direction is effectively equivalent to a long-range interaction between two spins with distance $l = n+1$, which represents the entanglements of all spins in a plane. This equivalence is validated for every interaction along the third direction in the 3D Ising system. In the 2D Ising model with a magnetic field, the magnetic field is effectively equivalent to an interaction between two spins with distance $l = 1, 2, 3, \ldots, n+1$. The effective interaction changes from the nearest neighboring interaction, to the short-range interaction, and then to the long-range interaction, while the number of the spins involved in entanglements is changing. This fact implies that the approaches developed for the 3D Ising model must be amended for solving the 2D Ising model with a magnetic field.

## 3. Clifford algebraic approach

### 3.1. Jordan-von Neumann-Wigner framework and time average

The Clifford algebraic approach developed for solving exactly the ferromagnetic 3D Ising model in zero magnetic field [10] can be modified to be appropriate for the 2D Ising model at a magnetic field. The deriving processes are described as follows:

In order to overcome the difficulties caused by the non-commutation of operators, the system must be set up within the Jordan-von Neumann-Wigner framework [27] that forms the mathematical bases of the quantum mechanics and the quantum statistical mechanics. By utilizing the Jordan algebra for two operators,

$$\Gamma_i \circ \Gamma_j = \frac{1}{2}\left(\Gamma_i \Gamma_j + \Gamma_j \Gamma_i\right) \tag{21}$$

one can demonstrate that the generalization of the Jordan algebra for *n* operators equals to the average over the sum of $2^{n-1}$ terms of products of these operators. Namely, we have [10]:

$$\Gamma_1 \circ \Gamma_2 \circ \Gamma_3 \circ \Gamma_4 \cdots \circ \Gamma_{n-1} \circ \Gamma_n = \frac{1}{2^{n-1}} \sum_{t=1}^{2^{n-1}} P_t[\Gamma_1 \Gamma_2 \Gamma_3 \Gamma_4 \cdots \Gamma_{n-1} \Gamma_n] \tag{22}$$

where $P_t[\Gamma_1 \Gamma_2 \Gamma_3 \Gamma_4 \cdots \Gamma_{n-1} \Gamma_n]$ represents all the possible permutations of *n* operators. It is evident that such an average on the products of the operators is equivalent to a time average over all the possible states of the system. Note that in the original formulas in the statistical mechanics and the quantum statistical mechanics, it is necessary to perform the time average. Readers refer to Theorem 4 (Commutation theorem) proven in pages 15-17 of ref. [10].

Introducing the third dimension for the time average expands the 2D Ising system to be (2+1) dimensional with a larger Hilbert space. The 2n-normalized eigenvectors $\Psi_{2D}$ (Eq. (54) of Kaufman [4]) are generalized to construct the 2nl-normalized eigenvectors $\Psi_{(2+1)D}$. The eigenvectors $\Psi_{(2+1)D}$ are represented as two of three rows in Eq. (33) of ref. [8], but at the present stage without the weight factors $w_x$ and $w_y$. By using some basic facts of the direct product and the trace, the procedure of introducing an additional dimension maintains the physical properties (such as the trace of the transfer matrices, the partition function, the free energy and the thermodynamic properties) of the system unchanged. Readers refer to Theorem 1 (Trace invariance theorem) proven in pages 6-8 of ref. [10].

3.2. Linearization process

We have to deal with the nonlinear terms in the transfer matrix $\boldsymbol{V_3}$. The internal factors in the transfer matrix can be expressed as:

$$W_j = \prod_{k=1}^{j-1} i\,\Gamma_{2k-1}\Gamma_{2k} \tag{23}$$

which has the character analogous to that of the boundary factor $U$, splitting the space of the transfer matrices and the Hilbert space, as the Kaufman's procedure for the 2D Ising model [4]. The internal factors $W_j$ can be taken as the projection operator $P \equiv \frac{1}{2}\left[1 \pm W_j\right]$, which projects the system from a larger Hilbert space with more number of states to many smaller Hilbert subspaces with less number of states. Linearizing the nonlinear terms in the problem is realized by the projection operators $P$. There are $2^n$ terms of combinations of positive/negative signs, representing $2^n$ subspaces. According to the largest eigenvalue principle [4,10], only the largest eigenvalue contributes dominantly to the partition function $Z$ of the 2D Ising model in the thermodynamic limit. This processes above overcome difficulties (such as nonlocality, nonlinearity, non-commutative and non-Gaussian) for solving the problem.

### 3.3. Topological Lorentz transformation

Meanwhile, we have to account for the contributions of the nonlinear terms to the partition function and subsequent thermodynamic properties. According to the topology theory, the crosses of knots/links in the topological structures contribute also to the partition function of a physical system [28]. There is a mapping between a cross and a spin, and in the 2D Ising model at the magnetic field two contributions consist of local spin alignments and nonlocal spin entanglements. In Figure 4 of [11] and Figure 3 of [12], the nonlinear terms in the transfer matrices of the 3D Ising model were illustrated as lines of knots with crossings. The nontrivial topological structures with knots represent the long-range spin entanglements, which are equivalent to spin chains (see Figure 5 of [12]), which can be trivialized in higher dimensions (Figures 7 and 8 of [12]). The nontrivial topological structures in the 2D Ising model at the magnetic field are similar to those in the 3D Ising model. The only difference is that at the present system the number of the spins in the spin chains at different lattice points varies, while the number of the crossings in the knots changes with changing the lattice point. Since the nontrivial topological structures in the present system are analogous to the 3D Ising model, the topological Lorentz transformation developed for the 3D Ising model [10] can be modified to be appropriate for the present system. The topological Lorentz transformation for the 2D Ising model at the magnetic field can be described as follows:

According to Onsager [3] and Kaufman [4], the planar rotations in the spinor representation can be transformed into the rotation representation. The eigenvalues of the partition function can be calculated by the planar rotations in the rotation representation. The transfer matrices $\boldsymbol{V_1}$ and $\boldsymbol{V_2}$ in the spinor representation can be transformed to the matrices $\boldsymbol{V_0^-}$ (Eq. (41) in [4]) with products of the rotation matrices as represented in Eqs. (42) and (43) of [4]. Here for simplicity, we only represent the basic block element in these 2n-dimensional rotation matrices. The 2-dimensional block for the first (and last) product representing the rotation is expressed as:

$$r_1 = \begin{bmatrix} coshK_1^* & isinhK_1^* \\ -isinhK_1^* & coshK_1^* \end{bmatrix} \tag{24}$$

The 2-dimensional block in the middle product has the form:

$$r_2 = \begin{bmatrix} cosh2K_2 & isinh2K_2 \\ -isinh2K_2 & cosh2K_2 \end{bmatrix} \tag{25}$$

In the present system, the effects of the magnetic field add two rotation matrices in the middle product:

$$r_3 = \begin{bmatrix} cosh2H & isinh2H \\ -isinh2H & cosh2H \end{bmatrix} \tag{26}$$

and

$$r_4 = \begin{bmatrix} cosh2H' & isinh2H' \\ -isinh2H' & cosh2H' \end{bmatrix} \tag{27}$$

The former accounts for the local contribution, while the latter corresponds to a Lorentz transformation for the nonlocal contribution.

The rotation transformation matrix $\boldsymbol{R_0^-}$ for the 2D Ising model at a magnetic field could be written schematically as [4]:

$$
\boldsymbol{R_0^-} = \begin{bmatrix} \boldsymbol{a} & \boldsymbol{b} & 0 & 0 & 0 & \cdots & 0 & \boldsymbol{b}^* \\ \boldsymbol{b}^* & \boldsymbol{a} & \boldsymbol{b} & 0 & 0 & \cdots & 0 & 0 \\ 0 & \boldsymbol{b}^* & \boldsymbol{a} & \boldsymbol{b} & 0 & \cdots & 0 & 0 \\ 0 & 0 & \boldsymbol{b}^* & \boldsymbol{a} & \boldsymbol{b} & \cdots & 0 & 0 \\ & & & & & & & \\ & & & & & & & \\ \boldsymbol{b} & 0 & 0 & 0 & 0 & \cdots & \boldsymbol{b}^* & \boldsymbol{a} \end{bmatrix}
\tag{28}
$$

where

$$
\boldsymbol{a} = \begin{bmatrix} cosh2(K_2 + H + H') \cdot cosh2K_1^* & -icosh2(K_2 + H + H') \cdot sinh2K_1^* \\ icosh2(K_2 + H + H') \cdot sinh2K_1^* & cosh2(K_2 + H + H') \cdot cosh2K_1^* \end{bmatrix}
\tag{29}
$$

$$
\boldsymbol{b} = \begin{bmatrix} -\frac{1}{2} sinh2(K_2 + H + H') \cdot sinh2K_1^* & isinh2(K_2 + H + H') \cdot sinh^2K_1^* \\ -isinh2(K_2 + H + H') \cdot cosh^2K_1^* & -\frac{1}{2} sinh2(K_2 + H + H') \cdot sinh2K_1^* \end{bmatrix}
\tag{30}
$$

$$
\boldsymbol{b}^* = \begin{bmatrix} -\frac{1}{2} sinh2(K_2 + H + H') \cdot sinh2K_1^* & isinh2(K_2 + H + H') \cdot cosh^2K_1^* \\ -isinh2(K_2 + H + H') \cdot sinh^2K_1^* & -\frac{1}{2} sinh2(K_2 + H + H') \cdot sinh2K_1^* \end{bmatrix}
\tag{31}
$$

The topological Lorentz transformation (Eq. (27)) also serves as a local gauge transformation, while generating topological phases on eigenvalues and eigenvectors. The rotation in the third dimension stands for the topological contributions to physical properties. Meanwhile, the topological phases emerge on the eigenvectors $\Psi_{(2+1)D}$ so that the eigenvectors become $\Psi'_{(2+1)D}$, as represented as two of three rows in Eq. (33)

of ref. [8] with the weight factors $w_x$ and $w_y$ that can be replaced by $|w_x|\cos\phi_x$ and $|w_y|\cos\phi_y$, respectively, with $|w_x| = |w_y| \equiv 1$ [9]. The 2D Ising model at a magnetic field can be transferred by the topological Lorentz transformation from a nontrivial topological basis to a trivial topological basis, while taking into account the contribution of the nontrivial topological structures to the partition function and the thermodynamic properties. Finally, the desired solution is realized for the 2D Ising model at a magnetic field by fixing the rotation angle for the topological Lorentz transformation (*i.e.,* the local gauge transformation) and the phase factors. The generalized Yang-Baxter equation (so-called tetrahedron equation) guarantees the integrability of the present 2D Ising model at a magnetic field, via the topological transformation keeping the invariant of bracket polynomial (partition function) under the Reidemeister moves II and III [9-12]. The rotation angle is determined by the star-triangle relation (*i.e,* Yang-Baxter equation) and by an average process. The rotation angle for the topological Lorentz transformation and the topological phases are investigated in subsections 5.2 and 5.3, respectively.

## 4. Eigenvalues, partition function and magnetization

### 4.1. Eigenvalues and partition function

Then the procedure for solving exactly the 2D Ising model at a magnetic field is straightforward following the procedure developed by Onsager [3] and Kaufman [4]. The 2n-eigenvalues of the rotation transformation matrix $\boldsymbol{R}_{\mathbf{0}}^{-}$ are the eigenvalues of the $n$ 2-dimensional matrices $\boldsymbol{\alpha}_{2j} = \boldsymbol{a} + \epsilon^{2j} \cdot \boldsymbol{b} + \epsilon^{2(n-1)j} \cdot \boldsymbol{b}^{*}$ with $\epsilon \equiv e^{\frac{i\pi}{n}}$ [4]. The determinant of this rotation transformation matrix is +1. The eigenvalues of the 2D

Ising model at a magnetic field is expressed as:

$$cosh\gamma_{2t} = cosh2K_1^*cosh2(K_2 + H + H') - sinh2K_1^*sinh2(K_2 + H + H') \times \left[cos\omega_x cos\phi_x + cos\omega_y cos\phi_y\right] \quad (32)$$

where $H' = \frac{K_2 H}{2K_1}$ (see subsection 5.2 for a detail description). The geometric relationships for a hyperbolic triangle are similar to those of the 2D Ising model at zero magnetic field [3,4], which are represented in the Poincaré disk model with some modifications.

The partition function of the 2D Ising model at a magnetic field is represented as:

$$\mathrm{N}^{-1}\ln \mathrm{Z} = \ln 2 + \frac{1}{2(2\pi)^3}\int_{-\pi}^{\pi}\int_{-\pi}^{\pi}\int_{-\pi}^{\pi}\ln[cosh2K_1cosh2(K_2 + H + H') - sinh2K_1cos\omega' - sinh2(K_2 + H + H') \times \left[cos\omega_x cos\phi_x + cos\omega_y cos\phi_y\right]\Big] \mathrm{d}\omega'\mathrm{d}\omega_\mathrm{x}\mathrm{d}\omega_\mathrm{y} \quad (33)$$

The topological phases $\phi_x$ and $\phi_y$ at finite temperatures equal to $2\pi$ and $\pi/2$, respectively. In Eqs. (32) and (33), when $H = 0$, one has $H' = 0$, the solutions return to the 2D Ising model at the absence of a magnetic field. From the partition function, we can calculate the free energy and the specific heat, and obtain the critical exponent $\alpha = 0$ for the specific heat.

At the beginning of the procedure, we have pre-set the conditions $J_1 \geq J_2$. The eigenvalues (Eq.(32)) and the partition function (Eq. (33)) are held only for such conditions. For $J_1 \leq J_2$, the interaction parameters $K_1$ and $K_2$ should be permutated.

4.2. Magnetization

The spontaneous magnetization of the 2D Ising magnet was calculated exactly by Yang using a perturbation procedure [5]. This technique was generated to derive the spontaneous magnetization of the 3D Ising model [8]. For the present case, if we treat the magnetic field $H$ as an interaction along the third crystallographic direction, the model is transformed into the AMC model of the 3D Ising model. Then, as a weak virtual magnetic field ℵ is introduced, the perturbation procedure can be employed to calculate the magnetization of the 2D Ising model at a magnetic field $H$. The procedure is straightforward with replacements of parameters.

The magnetization of the 2D Ising model at a magnetic field is represented as:

$$M = \left[1 - \frac{16x_1^2 x_2^2 x_3^2 x_4^2}{(1 - x_1^2)^2 (1 - x_2^2 x_3^2 x_4^2)^2}\right]^{\frac{1}{8}} \tag{34}$$

with $x_1 = e^{-2K_1}$, $x_2 = e^{-2K_2}$, $x_3 = e^{-2H}$ and $x_4 = e^{-2H'}$. Note that the critical exponent β equals to 1/8, since the magnetic field does not alter the dimensionality of the 2D Ising model. Again, the magnetization (Eq.(34)) is validated only for the conditions $J_1 \geq J_2$. For $J_1 \leq J_2$, the interaction parameters $K_1$ and $K_2$ should be permutated.

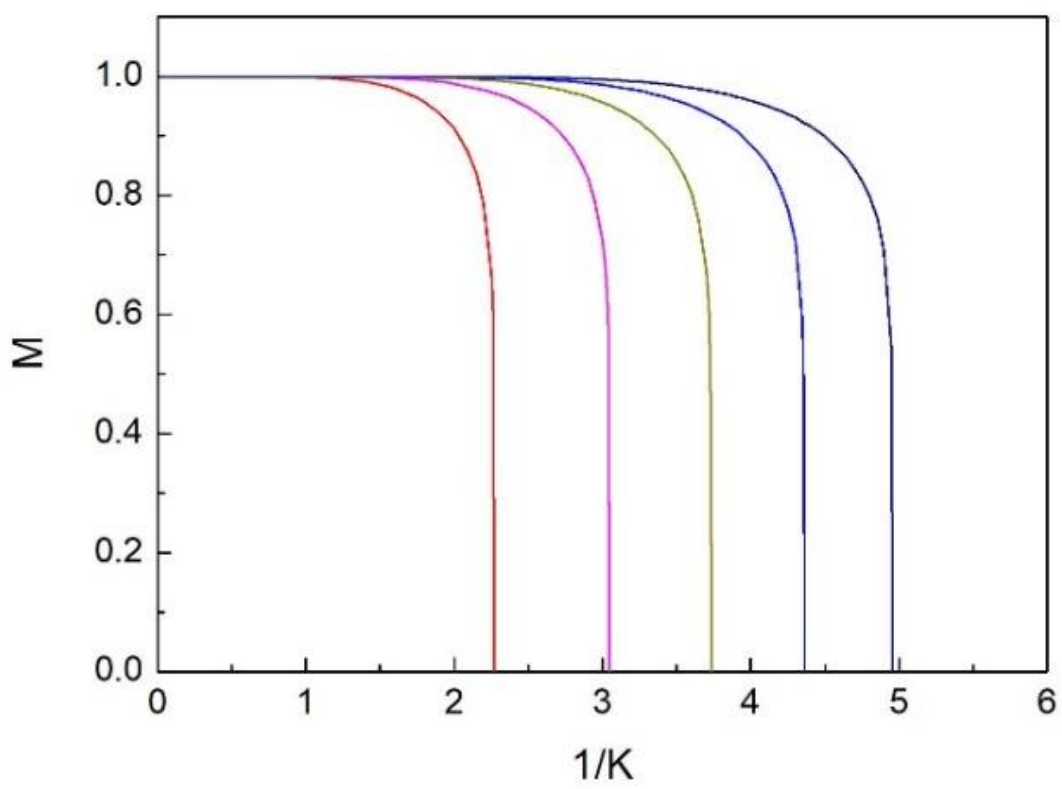

Figure 4. Temperature dependence of the magnetization at a magnetic field for the 2D Ising model with $K_1 = K_2 = K$. The curves from left to right correspond to the magnetic field $H$ = 0, 0.5$K$, $K$, 1.5$K$ and 2$K$, respectively.

Figure 4 represents the temperature dependence of the magnetization at a magnetic field for the square Ising model with $K_1 = K_2 = K$. At zero magnetic field, the magnetization agrees with Yang's exact solution for the spontaneous magnetization [5]. The critical point for the ferromagnetic – paramagnetic phase transition is located at the silver point, $x_c = e^{-2K_c} = \sqrt{2} - 1$, $\frac{1}{K_c} = 2.26918531 \ldots$ [2-5]. Application of a magnetic field increases the magnetization, shifting the critical point to higher temperatures ($\frac{1}{K_c} \approx 4.9564$ for $H$ = 2$K$). It is understandable that the magnetic field acts as an effective interaction to maintain the ordering of spins to suppress the thermal activation energy.

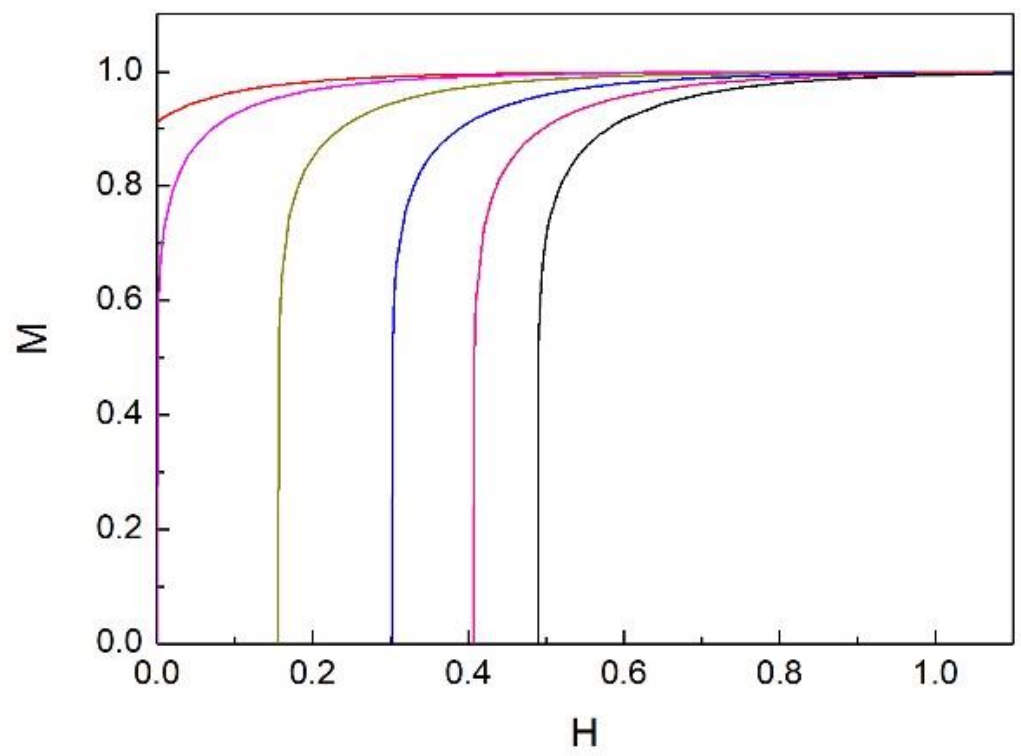

Figure 5. Magnetization processes at a fixed temperature for the 2D Ising model with $K_1 = K_2 = K$. The curves from left to right correspond to the temperature $1/K$ = 2, 2.269185…, 3, 4, 5 and 6, , respectively.

Figure 5 illustrates the magnetization processes at a fixed temperature for the 2D Ising model on a square lattice with $K_1 = K_2 = K$. At the temperature below the critical point, the magnetization increases continuously and gradually from the spontaneous magnetization to the saturation magnetization. At the critical point $\frac{1}{K_c} =$ 2.26918531 ..., the magnetization increases continuously and gradually from zero to the saturation. At the temperature above the critical point, the magnetization keeps zero unchanged until a critical field at which the magnetization jumps rapidly. The magnetic field together with the interactions balances with the thermal activation energy to maintain the paramagnetic state. At the critical field, such a balance is broken so that such a jump occurs. Then the magnetization increases gradually with increasing the magnetic field until the saturation. The exact solution for the magnetization of the 2D Ising model at a magnetic field reveals the competition between the magnetic field, the interactions and the temperature. In the system, the former two factors prefer the ordering state, whereas the latter factor favors the disordering state.

It is important to compare with experiments and to understand in-depth the jump in magnetization above the critical point as revealed in Figure 5. In experiments, the jump at the nonzero magnetic field above the critical point has been rarely reported in the magnetization curve $M \sim h$, but a jump often appears in the curves of $M^{1/\beta} \sim \left(\frac{h}{M}\right)^{1/\gamma}$ plotted for determinations of the critical exponents β and γ. Notice that $\left(\frac{h}{M}\right)^{1/\gamma} = \left(\frac{1}{\chi}\right)^{1/\gamma} = (T - T_c) = \Delta T$. The key here is to renormalize the magnetic field $H$ with respect to the relative temperature $\Delta T$. The critical point $T_c$ can be viewed as the starting point of the paramagnetic phase, followed the Curie-Weiss law. The relative temperature $\Delta T$ is usually used for description of the critical behaviors. Above the critical point with a relative temperature $\Delta \mathrm{T}$, the magnetic field should have a corresponding increase $\Delta \mathrm{H} = H_c$ in order to turn the magnetization deviating from zero. When the magnetic field is renormalized by the critical field $H_c$ with respect to the relative temperature $\Delta \mathrm{T}$, the magnetization processes above the critical point shown in Figure 6 have the same character as that in the critical point. All the magnetization curves above the critical point almost coincide with that at the critical point in the low field region with a continuous increase of the magnetization from zero, and only have slight differences in the high field region. Therefore, the nature of such a jump in the nonzero magnetic field is neither a first-order magnetization process nor a phase transition. When the magnetic field is renormalized by the magnetization $M$ to be $H_M = HM$, the magnetization processes in Figure 7 have the same tendency of the experimental data for $M \sim h$. Nevertheless, the results in Figure 5 can be utilized to evaluate the competitions between the interactions, the magnetic field and the thermal activation.

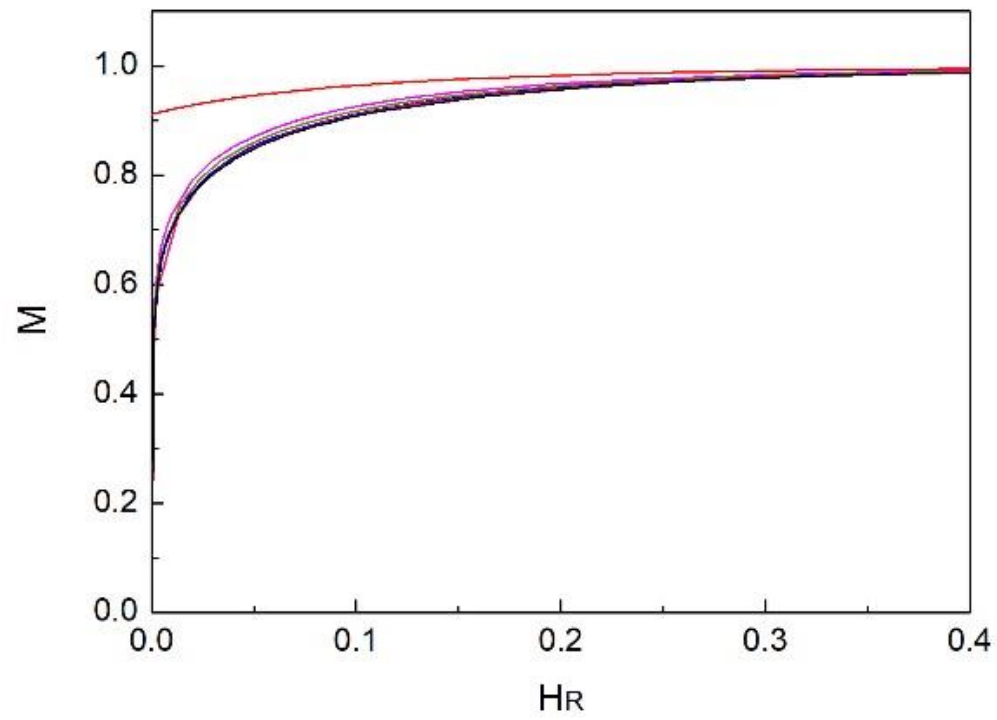

Figure 6. Magnetization processes at a fixed temperature for the 2D Ising model with $K_1 = K_2 = K$. The magnetic field is renormalized to be $H_R = H - H_c$. The curves from left to right correspond to the temperature $1/K$ = 2， 2.269185…, 3, 4, 5 and 6, , respectively.

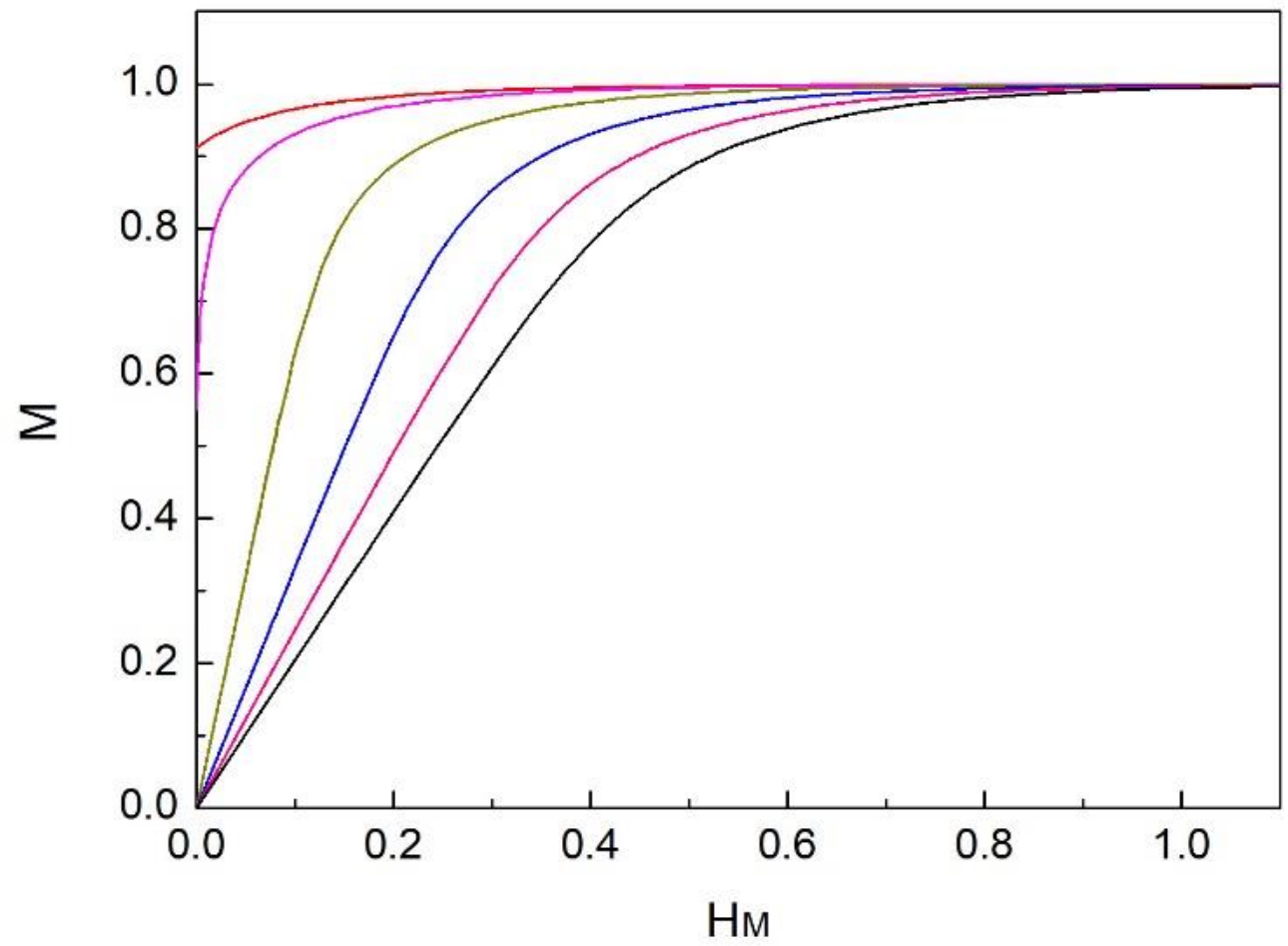

Figure 7. Magnetization processes at a fixed temperature for the 2D Ising model with $K_1 = K_2 = K$. The magnetic field is de-normalized to be $H_M$ = HM. The curves from left to right correspond to the temperature $1/K$ = 2， 2.269185…, 3, 4, 5 and 6, , respectively.

#### 4. Dimensionality, rotation angles and topological phases

4.1. Dimensionality

It is important to clarify the dimensionality of the 2D Ising model with a magnetic field. At first, it is emphasized that the nature of the system is of 2D characters. The topological structures are analogous to the AMC model for the 3D Ising models, which in a sense like 3D characters. In order to overcome the topological problems (including nonlocality, nonlinearity, non-commutative and non-Gaussian), the modified Clifford algebraic approach must be performed within a (2+1)-dimensional framework with an additional rotation for the topological Lorentz transformation (*i.e,* a gauge transformation). Utilizing Jordan algebras within the Jordan-von Neumann-Wigner framework [27], one also carries out a time average to form a new basis for eigenvectors and eigenvalues. The partition function expressed in Eq. (33) for the 2D Ising model with a magnetic field is a triple integral with two topological phases. As a whole, the 2D Ising model with a magnetic field can be treated as a quasi-2D system, because the magnetic field cannot be treated as a real interaction along the third crystallographic direction. As analyzed above, the 2D Ising model with a magnetic field can be seen as an AMC model being a 2D Ising model interacting with its nearest neighboring plane. It is equivalent to a two-level grid Ising model subtracts a 2D Ising model, namely, no intralayer interactions exist on the second layer of the two-level grid Ising model [23]. Evidently, the magnetic field cannot construct a full 3D lattice. Therefore, the dimensionality of the 2D Ising model keeps unchanged with the application of a magnetic field, which maintains the critical exponents ($\alpha = 0$ and $\beta = 1/8$) to be the 2D Ising universality.

4.2. Rotation angles

It is important to clarify the rotation angles for the transformations in the 2D Ising model with a magnetic field. The contributions of the nontrivial topological structures to the physical properties can be taken into account by the topological Lorentz transformation (*i.e.*, the gauge transformation), smoothing the crosses of knots/links. The Kramers-Wannier duality [2] is called the star-triangle relation corresponding to the Yang-Baxter equation [9]. The Yang-Baxter equation ensures the integrability of the system, while it represents Reidemeister moves guaranteeing the invariance of the physical properties during the trivialization process [9,10]. In the 3D case, the Yang-Baxter equation is generalized to be the generalized Yang-Baxter equation (so-called tetrahedron equation) [9,10], with its special solution being the star-triangle relation. From the star-triangle relation $KK^* = KK' + KK'' + K'K''$, the rotation angle $K''' = \frac{K'K''}{K}$ was determined for the 3D Ising model at zero magnetic field [8]. The star-triangle relation for the rotation angles of the 3D Ising model can used also for the 2D Ising model with a magnetic field, however, the modifications are needed. The modifications are caused by the difference between the nontrivial topological structures in the two models. In the 3D Ising model at zero magnetic field the rotation angle is unique for every local transformation at every Ising spin at the 3D lattice, because the effective long-range interactions act on two spins with the same distance $l = n+1$ and the spin entanglements are same for all the spin sites. In the 2D Ising model with a magnetic field, the rotation angle changes with altering the position of the spin through the 2D lattice, because the effective interaction changes from the nearest neighboring interaction to the long-range interaction, accompanying with changing the number of the spins involved in the entanglements. The rotation angle for the local gauge transformation should be proportional to the strength of the effective interaction replying on the distance of effectively interacted spins (i.e., the number of entangled spins). In the Clifford algebraic representation, the number of $\Gamma_{2j}$ matrices in the transfer matrix $\boldsymbol{V_3}$ increases as the odd numbers 1, 3, 5, …, 2*j*-1, with changing the lattice site *j* through the lattice. Such a change can be clearly seen from the schematic illustrations. For a spin at the left line of Figure 3, the nonlocal effect is strongest, which is the same as in the 3D case, so that the rotation angle is $H' = \frac{K_2 H}{K_1}$ as determined by the start-triangle relation (i.e., Yang-Baxter equation) $K_1 K_1^* = K_1 K_2 + K_1 H + K_2 H$. For a spin at the right line (next to a virtual boundary) in Figure 3, the magnetic field acts as a nearest neighboring interaction without the nonlocal effect, so that the rotation angle is zero. Therefore, the final results equal to the average of the rotation angles through the 2D lattice. We have the rotation angle $H' = \frac{K_2 H}{2K_1}$ for the 2D Ising model with a magnetic field.

4.3. Topological phases

It is important to clarify the topological phases. The topological phases generate on eigenvectors (and eigenvalues) of the 2D Ising model with a magnetic field. The singularities are caused by crossings in the nontrivial topological structures. To trivialize the nontrivial topological structures, one needs to carry out a gauge transformation (i.e., the topological Lorentz transformation). The Röhrl Theorem [11,29] provides the possibility of the existence of a multi-valued function with regular singularities for a given monodromy representation. The topological phases can be described also by the Gauss-Bonnet-Chern formula [12,30], which are significances analogous to the phase factors in the Aharonov-Bohm effect [31], the Berry phase effect [32], *etc.* Notice that for the 3D Ising model at zero magnetic field, there are three topological phases, while for the 2D Ising model with a magnetic field, there are two topological phases.

## 5.Conclusions

In conclusion, the exact solution of the rectangular Ising model at an external magnetic field is derived by a modified Clifford algebraic approach. Inspecting on the transfer matrices by the Clifford algebraic representation, the transfer tensor representation and the schematic representation confirms the existence of the nontrivial topological structures in this system. The similarity and the difference between the nontrivial topological structures in the 2D Ising model with a magnetic field and the 3D Ising model at zero magnetic field are clarified. A topological Lorentz transformation is applied for dealing with the topological problem in the present system. The rotation angle for the transformation is determined by the Yang-Baxter relations and subsequently by averaging the rotation angles that are proportional to the topological actions of spins at the 2D lattice. The eigenvalues, the partition function and the magnetization are obtained analytically. Application of a magnetic field increases the magnetization, shifting the critical point to higher temperatures. At the temperature above the critical point, the magnetization keeps zero unchanged a critical field $H_c$ at which the magnetization jumps rapidly. The appearance of such a jump in the magnetization reveals the competitions between the interactions, the magnetic field and the thermal activation. When the magnetic field is renormalized by the critical field with respect to the relative temperature $\Delta\mathrm{T} = (T - T_c)$, the magnetization processes above the critical point have the same character as that in the critical point. The nature of such a jump in the nonzero magnetic field is neither a first-order magnetization process nor a phase transition. The jump in the magnetization disappear if the magnetic field is renormalized to be $H_M = HM$, consistent with the experimental data for $M \sim h$. The dimensionality, rotation angles and topological phases are discussed.

Solving exactly the Ising models not only understands in-depth these models themselves [8-15], but also benefit to solving the hard problems in mathematics and computer sciences [20-23,33,34].

**Acknowledgements**

This work has been supported by the National Natural Science Foundation of China under grant number 52031014.

**Data availability** Data available upon request from the author.

**Conflict of interest** The author declares that this contribution is no conflict of interest.

**Author contribution statements** Z.D. Zhang is the only author, who contribute to conception, method, investigation, validation, visibility and writing the manuscript.

**References**

[1] E. Ising, Beitrag zur Theorie des Ferromagnetismus, Z Phys. **31**, 253-258 (1925).

[2] H.A. Kramers, G.H. Wannier. Statistics of the two-dimensional ferromagnet, Phys. Rev. **60** (3), 252-262 (1941).

[3] L. Onsager, Crystal statistics I: A two-dimensional model with an order-disorder transition, Phys. Rev. **65** (3-4), 117-149 (1944).

[4] B. Kaufman, Crystal statistics II: Partition function evaluated by spinor analysis, Phys. Rev. **76** (8), 1232-1243 (1949).

[5] C.N. Yang, The spontaneous magnetization of a two- dimensional Ising model, Phys. Rev. **85** (5), 808-816 (1952).

[6] M. Kac, J.C. Ward, A combinatorial solution of the 2-dimensional Ising model, Phys. Rev. **88** (6), 1332-1337 (1952).

[7] G.F. Newell, E.W. Montroll, On the theory of the Ising model of ferromagnetism, Rev. Mod. Phys. **25** (2), 353-389 (1953).

[8] Z.D. Zhang, Conjectures on the exact solution of three - dimensional (3D) simple orthorhombic Ising lattices, Phil. Mag. **87** (34), 5309-5419 (2007).

[9] Z.D. Zhang, Mathematical structure of the three-dimensional (3D) Ising model, Chinese Phys. B **22** (3), 030513 (2013).

[10] Z.D. Zhang, O. Suzuki, N.H. March, Clifford algebra approach of 3D Ising model, Advances in Applied Clifford Algebras **29** (1), 12 (2019).

[11] O. Suzuki, Z.D. Zhang, A method of Riemann-Hilbert problem for Zhang's conjecture 1 in a ferromagnetic 3D Ising model: trivialization of topological structure, Mathematics **9** (7), 776 (2021).

[12] Z.D. Zhang, O. Suzuki, A method of the Riemann-Hilbert problem for Zhang's conjecture 2 in a ferromagnetic 3D Ising model: topological phases, Mathematics **9** (22), 2936 (2021).

[13] Z.D. Zhang, Exact solution of two-dimensional (2D) Ising model with a transverse field: a low-dimensional quantum spin system, Physica E **128**, 114632 (2021).

[14] Z.D. Zhang, Exact solution of the three-dimensional (3D) $Z_2$ lattice gauge theory, Open Physics **23**, 20250215 (2025).

[15] Z.D. Zhang, Topological quantum statistical mechanics and topological quantum field theories, Symmetry **14** (2), 323 (2022).

[16] Z.D. Zhang, N.H. March, Three-dimensional (3D) Ising universality in magnets and critical indices at fluid-fluid phase transition, Phase Transitions **84** (4), 299-307

(2011).

[17] Z.D. Zhang, Universality of critical behaviors in the three-dimensional (3D) Ising magnets, arXiv: 2510.09111.

[18] B.C. Li, W. Wang, Exploration of dynamic phase transition of 3D Ising model with a new long-range interaction by using the Monte Carlo Method, Chinese Journal of Physics **90**, 15-30 (2024).

[19] B.C. Li, W. Wang, Influence of a new long-range interaction on the magnetic properties of a 2D Ising layered model by using Monte Carlo method, Chinese Journal of Physics **87**, 525-539 (2024).

[20] Z.D. Zhang, Computational complexity of spin-glass three-dimensional (3D) Ising model, J. Mater. Sci. Tech. **44**, 116-120 (2020).

[21] Z.D. Zhang, Mapping between spin-glass three-dimensional (3D) Ising model and Boolean satisfiability problems, Mathematics **11** (1), 237 (2023).

[22] Z.D. Zhang, Lower bound of computational complexity of Knapsack problems, AIMS Math. **10** (5), 11918-11938 (2025).

[23] Z.D. Zhang, Relationship between spin-glass three-dimensional (3D) Ising model and traveling salesman problems, arXiv: 2507.01914.

[24] Z.D. Zhang, Two-dimensional Ising model with the next nearest interactions, Phys. Rev. E 113 (2026) 065111.

[25] P. Jordan and E. Wigner, Über das Paulische Äquivalenzverbot, Z. Phys. **47**, 631 631-651 (1928).

[26] J.H.H. Perk, Comment on 'Conjectures on exact solution of three-dimensional (3D)

simple orthorhombic Ising lattices', Phil. Mag. **89**, 761-764 (2009).

[27] P. Jordan, J. von Neumann, E. Wigner. On an algebraic generalization of the quantum mechanical formalism, Ann. of Math. **35**, 29-64 (1934).

[28] L.H. Kauffman, The mathematics and physics of knots, Rep. Prog. Phys. **68** (12), 2829-2857 (2005).

[29] H. Röhrl, Das Riemannsch-Hilbertsche problem der theorie der linieren differentialgleichungen, Math. Ann. **133**, 1-25 (1957).

[30] S.S. Chern, On the curvatura integra in a Riemannian manifold, Ann. Math. **46**, 674-684 (1945).

[31] Y. Aharonov, D. Bohm, Significance of electromagnetic potentials in the quantum theory, Phys. Rev. **115** (3), 485-491 (1959).

[32] M.V. Berry, Quantal phase-factors accompanying adiabatic changes, Proc. R, Soc. London A **392** (1802), 45-57 (1984).

[33] Z.D. Zhang, Equivalence between the zero distributions of the Riemann zeta function and a two-dimensional Ising model with randomly distributed competing interactions, arXiv: 2411.16777.

[34] Z.D. Zhang, Equivalence between the pair correlation functions of primes and of spins in a two-dimensional Ising model with randomly distributed competing interactions, arXiv: 2508.14938.